\documentclass[11pt,a4paper]{article}
\pdfoutput=1
\usepackage{a4wide}
\usepackage{amsfonts,amssymb,amsmath}
\usepackage{graphicx}
\usepackage{booktabs}
\usepackage{array}
\usepackage{color}
\usepackage[usenames,dvipsnames]{xcolor}
\usepackage{ulem}
\usepackage[small,bf]{caption}
\usepackage{cite}
\usepackage{verbatim}
\usepackage{dsfont}
\usepackage{hyperref}

\allowdisplaybreaks

\numberwithin{equation}{section}

\begin{document}
\begin{titlepage}

\vspace*{-15mm}
\begin{flushright}
NCTS-PH/1702
\end{flushright}
\vspace*{0.7cm}

\begin{center}
{
\bf\LARGE
Detection prospects for the Cosmic Neutrino \\[0.2em]
Background using laser interferometers
}
\\[8mm]
Valerie~Domcke$^{\, a,}$ \footnote{E-mail: \texttt{valerie.domcke@apc.univ-paris7.fr}} and
Martin~Spinrath$^{\, b,}$ \footnote{E-mail: \texttt{martin.spinrath@cts.nthu.edu.tw}},
\\[1mm]
\end{center}
\vspace*{0.2cm}
\centerline{$^{a}$ \it AstroParticule et Cosmologie (APC)/Paris Centre for Cosmological Physics,}
\centerline{\it Universit\'{e} Paris Diderot}
\vspace*{0.2cm}
\centerline{$^{b}$ \it Physics Division, National Center for Theoretical Sciences,}
\centerline{\it National Tsing-Hua University, Hsinchu, 30013, Taiwan}
\vspace*{1.20cm}

\begin{abstract}
\noindent
The cosmic neutrino background is a key
prediction of Big Bang cosmology which has not been
observed yet. 
The movement of the earth through this neutrino bath
creates a force on a pendulum,
as if it were exposed to a cosmic wind.
We revise here estimates for the resulting pendulum acceleration
and compare it to the theoretical sensitivity of an experimental
setup where the pendulum position is measured using current
laser interferometer technology as employed in gravitational wave
detectors. We discuss how a significant improvement of this setup
can be envisaged in a micro gravity environment. The proposed setup
could also function as a dark matter detector in the sub-MeV
range, which currently eludes direct detection constraints.
\end{abstract}

\end{titlepage}
\setcounter{footnote}{0}

\section{Introduction}

The cosmic neutrino background (CNB) is a robust prediction of the standard
model of particle physics in standard $\Lambda$CDM cosmology. A measurement
of this elusive background would confirm or challenge our understanding of
these standard models up to an energy scale of about 1~MeV, far beyond the
reach of its cousin, the cosmic microwave background (CMB), which over the
last decades has provided us with invaluable information covering
cosmological energy scales up to about 0.3~eV. Although very
abundant in the universe, the weakly interacting nature of the cosmic
neutrinos (which makes them so valuable to probe the early universe) makes
them inherently hard to detect. Several proposals have been put forward,
for some recent overviews,  see, e.g., \cite{Ringwald:2009bg, Vogel:2015vfa},
all of them beyond the reach of current technology.
The most promising proposal at the moment seems to be the PTOLEMY experiment
\cite{Betts:2013uya} which aims at detecting the CNB through the inverse beta
decay of tritium. However, very recently, the first discovery of gravitational
waves by the LIGO/VIRGO collaboration~\cite{Abbott:2016blz} proved the
possibility of detecting an even more elusive potential messenger of our cosmic
past, namely gravitational waves. In this paper, we investigate the prospects of
searching for the CNB with laser interferometer technology, similar to the technology
currently developed for gravitational wave detectors. A related proposal was
recently put forward for searching for sub-eV dark matter~\cite{Graham:2015ifn}.

Let us briefly recall some key features of the CNB.
As the temperature decreases in the course of the evolution of
the Universe, the weak interactions keeping neutrinos in equilibrium
with the thermal bath freeze out (at about $T \sim 1$~MeV) and
the CNB decouples from the thermal bath of photons, electrons and positrons.
At $T \sim 0.5$~MeV, the production of electrons and positrons freezes out,
leading to a reheating of the photon bath sourced by the annihilation of
electrons and positrons. Consequently, the CNB temperature $T_\nu$ is predicted
to be slightly lower than the observed CMB temperature of $T_0=2.735$~K,
$T_\nu = (4/11)^{1/3} \,T_0 \approx 1.95$~K, which corresponds to a
thermal energy of $k_B T_\nu \approx 0.16$~meV. The average neutrino number
density today is determined by the thermal neutrino abundance at the time of
decoupling, $\bar n_\nu = 3/22 \, n_\gamma \simeq 56\, \text{cm}^{-3}$ per flavour
and per chirality.\footnote{
Recently, it was also discussed that the CNB might have a non-thermally
produced component \cite{Chen:2015dka} increasing the total neutrino number
density by an ${\cal O}(1)$ factor. Since the momentum distribution of this
non-thermal component is model-dependent, we will not address this
possibility here any further.
}
This average density as well as the momentum distribution may
however be altered locally if the neutrinos are heavy enough to cluster to
astrophysical gravitational structures such as our galaxy.

The remainder of this paper is organized as follows. In Sec.~\ref{sec:review}
we review the predicted local density and momentum distribution of the CNB
neutrinos. In Sec.~\ref{sec:exp}, after a brief sketch of a possible simple
experimental setup,  we propose avenues how a significant improvement might
be achieved. 
Then we update and revise theoretical expectations for a mechanical acceleration
induced by the CNB wind in Sec.~\ref{sec:theory}. Comparing these values with
current interferometer technology as employed by gravitational wave experiments,
we conclude that in the simplest setup the sensitivity still falls short by many
orders of magnitude using current technology. For comparison we also
quote estimates for the solar neutrino wind and a possible dark matter wind.
In the latter case, the expected sensitivity is many orders of magnitude below
current direct detection bounds for dark matter masses above a few GeV, but could
provide competitive bounds for elastic dark matter - nucleon scattering in the sub-MeV
range.


\section{Neutrino masses and the CNB \label{sec:review}}

Contrary to the photons of the CMB, neutrinos have a (small) mass, rendering
the CNB phenomenology more diverse than the more familiar CMB.
Current bounds from laboratory experiments require
{$m_\nu \lesssim 2.8 $~eV/$c^2$~\cite{Weinheimer:1999tn}}, whereas cosmological
bounds are pushing
down to {$\sum m_\nu \lesssim 0.23 $~eV/$c^2$~\cite{Ade:2015xua}}. At
the same time, the heaviest eigenstate must be heavier than  about $0.05$~eV/$c^2$
to explain neutrino oscillation data~{\cite{PDG2016}}. Depending
on their mass, CNB neutrinos may be relativistic or non-relativistic 
and they may cluster gravitationally in the potential wells formed by dark
matter.
We will assume here that unclustered neutrinos have no average relative
velocity with respect to the CMB rest frame as measured by the CMB dipole.
Hence, 
for sufficiently light neutrinos the total neutrino flux on earth is simply the
average neutrino density,  $2 \cdot \bar n_\nu$ per flavour or mass eigenstate,
multiplied by the  velocity  $\beta^\text{CMB}_\oplus \,  c \approx 369$~km/s of
the earth traveling through the CNB rest frame as measured by the CMB dipole.
Since
the orientation of this dipole is known from the observation of the CMB, so is the
direction of the `neutrino wind' on earth. The momentum distribution of these
unclustered neutrinos is to good approximation a red-shifted copy of the Fermi-Dirac
distribution describing the neutrino bath at decoupling.

However, from neutrino oscillation data we know that at least two neutrino mass
eigenstates are non-relativistic,
$\sqrt{|\Delta m^2_{31}|} \, c^2 \gg \sqrt{|\Delta m^2_{21}|} \, c^2 \approx
8.5 \cdot 10^{-3}$~eV~$\gg 3.15 \, k_B \, T_\nu \approx 5 \cdot 10^{-4}$~eV. 
These non-relativistic neutrinos might gravitationally cluster to large DM
structures such as galaxies, clusters and superclusters. Clustering becomes relevant
if the intrinsic neutrino velocity drops below the escape velocity of the
corresponding astrophysical structure,
$v \sim \langle p_\nu \rangle c^2/ E_\nu \sim 3 \, k_B \, T_\nu \, c/m_\nu < v_\text{esc}$.
For the Milky way, the escape velocity is about $v_\text{esc}^\text{MW} \simeq 500$~km/s
while for our supercluster it is estimated to be ${\cal O}(10^3)$~km/s.
Gravitational clustering will enhance the local neutrino density and modify the momentum
distribution.
Ref.~\cite{Ringwald:2004np} studied the clustering of neutrinos to the Milky Way
and to supercluster structures, finding an enhancement factor of
$n_\nu/\bar{n}_\nu = {\cal O}(1 - 100)$,
depending on the mass of the neutrinos and the size of the astrophysical structure.

The velocity dispersion of these clustered neutrinos can be estimated from the virial
velocity, which for neutrinos bound to the galaxy is about $\beta_\text{vir} \, c \sim 10^{-3} \, c$
at the position of the earth. Up to an ${\cal O}(1)$ factor this agrees with the simulations
of Ref.~\cite{Ringwald:2004np}, which indicate that the momentum
distribution is well approximated by a Fermi-Dirac distribution with a cut-off around the
escape velocity. In the following we will take the velocity dispersion of clustered neutrinos
to be $\beta_\text{vir} \, c$. However, this value might be enhanced by a factor
of ${\cal O}(1 - 10)$.
Note that the direction of the neutrino wind will differ compared to the unbound case.
For unbound neutrinos, the neutrino wind is expected
at an angle of $\sim 10^\circ$ to the ecliptic plane, for bound neutrinos we
expect $\sim 60^\circ$~\cite{Safdi:2014rza}. In practice
the neutrino wind experienced on earth can be a combination of 
all these possibilities, including relativistic and non-relativistic states as well
as (partially) clustered populations.

In addition to the effects described above, gravitational focusing effects within the solar
system may induce an annual modulation of the neutrino rate on earth, which is
also sensitive to the neutrino mass~\cite{Safdi:2014rza}. It has also been argued that
the CNB could be asymmetric (i.e., containing different number densities of neutrinos and
anti-neutrinos) in one or more flavours, which could result in an
enhancement of the average neutrino density~\cite{Langacker:1982ih}. 
While Big Bang Nucleosynthesis severely constrains such an asymmetry
for electron neutrinos~\cite{Kang:1991xa,Lesgourgues:1999wu,Mangano:2011ip,Castorina:2012md}
the constraints on the muon and tau neutrinos (which would contribute to the extra
relativistic degrees of freedom as measured in the CMB) are much
weaker~\cite{Caramete:2013bua,Barenboim:2016shh,Barenboim:2017dfq}.

In the following we will hence distinguish three cases with increasing absolute neutrino
mass scale: relativistic (R), non-relativistic unclustered (NR-NC) and non-relativistic
clustered (NR-C) neutrinos. As a reference value we will work with the standard average
neutrino density per flavour or mass eigenstate set by
$2 \, \bar n_\nu = 112 \, \text{cm}^{-3}$.


\section{The experimental setup \label{sec:exp}}

In this section we will first discuss a simple toy setup
for the kind of experiment we are considering to estimate
the sensitivity which could be achieved
in the near future. A comparison with the magnitude of
the expected signal, derived in Sec.~\ref{sec:theory}, will reveal 
that current interferometer technology falls orders of magnitude
short of the sensitivty required to detect the CNB. With this
in mind, we limit our discussion in this section to a schematic
description of the possible experimental setup.
In Sec.~\ref{sec:future} we will outline some
potential modifications which might help to drastically increase
our expected sensitivity.

\subsection{Sketch of the experimental setup \label{sec:pendulum}}

\begin{figure}
  \centering
  \includegraphics[scale=0.8]{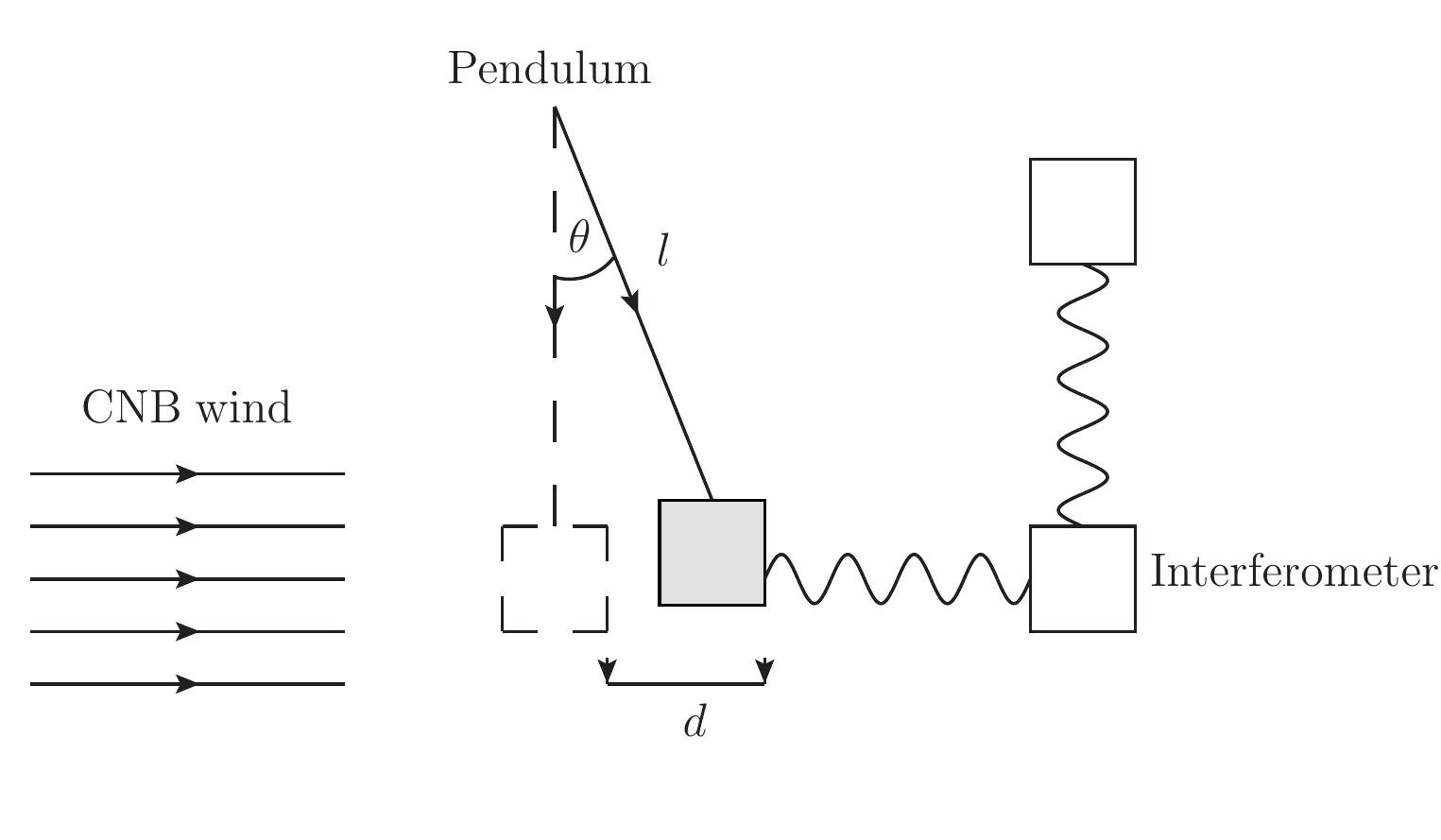}
  \caption{Sketch of the experimental setup. The neutrino wind exerts a force on the test mass at the end of the pendulum of
    length $l$. The resulting excursion, $d$ can be probed by the laser
    interferometer.
    }
  \label{fig:Setup}
\end{figure}

We consider test masses mounted on classical pendulums. The neutrino wind
will result in a force on the test masses which leads to an excursion in the direction
of the wind, see Fig.~\ref{fig:Setup}  If the force and the excursion $d$ are extremely
small and slowly
varying (see Sec.~\ref{sec:theory} for details), we may use the small angle approximation
\begin{equation}
 d = l \sin \theta \approx l \, \frac{a_\nu}{g} \;,
 \label{eq:excursion}
\end{equation}
where $l$ is the length of the pendulum and $g \approx 980$~cm/s$^2$ is the
standard acceleration due to the gravitational field of the earth.

Let us assume that the pendulum fits nicely into an ordinary lab and 
hence for simplicity we take $l = 100$~cm. As a reference value, the current
sensitivity of LIGO is $\Delta d = h \, L \, \Delta f^{1/2} \simeq 1 \cdot 10^{-17} (\Delta f / 10~\text{Hz})^{1/2}$~cm,
with $L = 4$~km denoting the length of the interferometer arms,
$h \simeq 10^{-23}/\sqrt{\text{Hz}}$ denoting the current peak strain sensitivity
at $f_0 = 100$~Hz and $\Delta f$ indicating the bandwidth used for the
analysis~\cite{TheLIGOScientific:2016agk}.
This translates to a sensitivity for the acceleration of
\begin{equation} \label{eq:sensitivity_now}
a_{\text{min}}^{\text{now}} = g \, \frac{\Delta d}{l} \approx 1 \cdot 10^{-16} \text{ cm/s}^2 \,,
\end{equation}
where we have set $\Delta f = 10$~Hz.
The design sensitivity of advanced LIGO is expected to lower this by a factor of 3, with
future upgrades expected to improve the current sensitivity by about a
factor 10~\cite{TheLIGOScientific:2016agk}. For the Einstein telescope with 10~km
arm length strain sensitivities of a few times $10^{-25}/\sqrt{\text{Hz}}$ are
envisaged~\cite{ET}. 
In the future one could thus optimistically estimate a sensitivity of 
\begin{equation} \label{eq:sensitivity}
a_{\text{min}} \approx  3 \cdot 10^{-18} \text{ cm/s}^2 \,,
\end{equation}
for terrestrial experiments.
Space based GW interferometers have been designed for
lower frequencies $f_0 \sim \text{few }$~mHz, but their expected
sensitivity in terms of absolute distance changes is lower
($\Delta d \simeq 10^{-11}$~cm for LISA~\cite{LISA} for a bandwidth of $\Delta f = 10^{-4}$~Hz).

Note, that we will assume here for simplicity that the
experiment has an optimal orientation with respect to the neutrino
wind, i.e., that the movement of the earth rotates
the experiment such that within a day, the pendulum interferometer
arm is orientated parallel as well as orthogonal to the neutrino wind. The optimal situation
is achieved for a neutrino wind orthogonal to the
earth's axis, in which case the pendulum interferometer arm can reach
an orientation parallel and anti-parallel to the neutrino wind within
a day. 
In a realistic setup there should also be an additional
annual modulation, see \cite{Safdi:2014rza} for a recent discussion.

It is crucial to note that our above estimates apply
for the frequency band of the corresponding detector. In particular, any terrestrial
setup will suffer drastically from seismic noise for frequencies below about 1~Hz.
On the contrary, the intrinsic frequency of a signal induced by the CNB is set by
the earth's rotation, 1/day $\sim 10^{-5}$~Hz. It thus seems extremely difficult at best to exploit
the remarkable sensitivity of laser interferometers to search for the CNB in an
earth based laboratory. A possible way out might be to focus not on the daily variation of
he excursion $d$, but on the high frequency component due to individual neutrino
interactions, governed by the rate of neutrinos scattering off the test mass. Optimizing the
setup for this measurement requires adjusting the pendulum length as well as the size and
material of the test mass. 

More concretely, using the expressions derived in Sec.~\ref{sec:theory}, we note
that the number of scattering events per unit time, $\Gamma$, depends on the mass of
the pendulum $M$ whereas the resulting acceleration $a_{G_F^2}$ to leading order does not,
\begin{equation}
a_{G_F^2} = R \, \langle \Delta p \rangle \,, \quad \Gamma = R \,  M \,.
\end{equation}
Here $\langle \Delta p \rangle$ denotes the average momentum transfer and $R$ denotes
the event rate per second and gram. For example, in the case
of non-relativistic non-clustered neutrinos and approximating LIGO's mirrors as
40~kg pure silicon, we find
\begin{equation}
\Gamma = 86~\text{Hz} \left( \frac{M}{40~\text{kg}} \right) \left( \frac{(A -  Z)^2/A^2 }{0.25} \right) \left( \frac{\rho }{2.34 \text{ g/cm}^3} \right) \,,
\end{equation}
which is right within the LIGO sensitivity band. However, as we will see in the
next section, the corresponding acceleration of $a_{G_F^2} = 7.3 \cdot 10^{-32}$~cm/$\text{s}^2$
for silicon is very far beyond the current reach of LIGO.
This acceleration could even drop down to  $a_{G_F^2} \approx 10^{-33}$~cm/$\text{s}^2$ for 
light normal hierarchical Majorana neutrinos

The situation is somewhat more optimistic for
DM searches, where the event rate is senstive to the mass and cross-section of the DM
particle. Using the same approximation as above to model LIGO's mirrors, we find that
LIGO's peak sensitivity of $100$~Hz corresponds approximately, e.g., to the following combinations of
acceleration, DM mass and cross-section:
\begin{align}
a_{G_F^2} & \approx 10^{-18}~\text{cm/s}^2 \,,  \quad  m_X = 10~\text{GeV} \,, \quad  \sigma_{X-N} = 3 \cdot 10^{-34}~\text{cm}^2  \,,\\
a_{G_F^2} & \approx 10^{-20}~\text{cm/s}^2 \,,  \quad  m_X = 0.1~\text{GeV} \,, \quad  \sigma_{X-N} = 3 \cdot 10^{-36}~\text{cm}^2  \,,\\
a_{G_F^2} & \approx 10^{-22}~\text{cm/s}^2 \,,  \quad  m_X = 1~\text{MeV} \,, \quad  \sigma_{X-N} = 1 \cdot 10^{-40}~\text{cm}^2  \,.
\end{align}
More details on the dependence of $\Gamma$ and $a_{G_F^2}$ on the DM mass and cross-section
are given in Sec.~\ref{sec:theory}.

The numbers found above have to be compared with Cavendish-type torsion balances which try to measure the
same kind of acceleration we study here \cite{Hagmann:1999kf}. Recent torsion-balance tests of the
weak equivalence principle have sensitivities for differential accelerations of the order
of $10^{-13}$~cm/s$^2$ \cite{Wagner:2012ui}, and it has been claimed that accelerations down to
$10^{-23}$~cm/s$^2$ may be reached \cite{Hagmann:1999kf}.
In \cite{Graham:2015ifn} it was argued that assuming an optimal, shot noise limited laser read-out,
torsion balance experiments can always do better than a linear displacement experiment as we
suggest here. To keep the discussion simple we will imagine a linear setup in the following,
keeping in mind that further improvements might be possible with different geometries.

As we will review in Sec.~\ref{sec:theory}, the expected displacements of the pendulum
due to the CNB are tiny, far beyond the sensitivity of current and upcoming laser interferometers. 
We hence do not want to make the discussion unnecessarily complicated with
lengthy musings about the exact shape of the modulation or the experimental setup.
With all these caveats in mind, we will 
nevertheless refer to Eqs.~\eqref{eq:sensitivity_now} and \eqref{eq:sensitivity}
as benchmark values for what might be achieved with this kind of experiment.
We do however dedicate the following subsection to some speculation on avenues
which might increase the sensitivity by many orders of magnitude.

\subsection{Possible avenues for significant improvement \label{sec:future}}

The two distinctive features of the signal 
are the directional information and the characteristic frequency.
Adding a second interferometer orthogonal to the first
to get two-dimensional information about the excursion of the pendulum or putting
several copies of the experimental setup on different locations on earth would help
to discriminate the signal from background. 

We further point out that the sensitivity
could be significantly improved if the acceleration $g$ is significantly lowered.
Implementing such a setup in space inside a rotating satellite
with tiny centrifugal forces might sound utopic now, but might
be possible in the future.
On the International Space Station routinely experiments
in micro gravity are performed with a net acceleration
of the order of $10^{-6}$~g \cite{Microgravity}.
Hence, the sensitivity of our setup placed in space
could conceivably be increased by six orders of magnitude or more.
This could also ameliorate the problem of
the required stability over time, since the rotation frequency of 
a space based experiment could be much faster than the corresponding
signal frequency of 1/day on earth.
One could also instead imagine to put the pendulum mass
in some kind of electromagnetic suspension to compensate
earth's gravity. Cleverly arranged, this might also damp
much larger background effects.

We also note the possibility of replacing the pendulum with two or more free falling
masses with different total neutrino cross sections - as we will see below
dramatically different cross sections can be obtained by varying the target
size and material due to an atomic enhancement factor. The free falling test
masses would thus drift apart under the influence of the neutrino wind,
seemingly violating the equivalence principle. In this context, it is remarkable
that LISA Pathfinder has probed the relative acceleration between two free
falling (identical) test masses down to $(5.2 \pm 0.1) \cdot 10^{-15} \, g/\sqrt{\text{Hz}}$
for frequencies around 1 mHz~\cite{Armano:2016bkm}.
Of course, a measurable effect would require a stability on much longer time scales.
A quick estimate shows that a cm - sized lead test mass subject to a constant
acceleration of $a = 10^{-27} \,\text{cm/s}^2$ over one month  by the CNB, see Sec.~\ref{sec:scattering},
 would be displaced by $a \, (30 \text{
days})^2/2 =  3 \cdot 10^{-15}$~cm. A distance which is in principle measurable with current laser technology.


\section{Theoretical expectations for the acceleration \label{sec:theory}}

The mechanical effect of the cosmic neutrino background has been known
already for a long time and we will refer here to the calculations of
Duda, Gelmini and Nussinov \cite{Duda:2001hd}. The formulas in this section
are mostly based on their work, subject to some improvements as we will detail below.
We restrict ourselves to the case of Dirac neutrinos which is the more
optimistic case for this kind of experiments.
For relativistic neutrinos the results
for Majorana and Dirac neutrinos are the same while for non-relativistic
neutrinos the $G_F$ effect would vanish for clustered neutrinos
and the $G_F^2$ effect is suppressed
by a factor of $(v_\nu/c)^2 \ll 1$ \cite{Duda:2001hd}.

\subsection{Magnetic torque (\texorpdfstring{$\boldsymbol{G_F}$}{GF} effect)}

There have been some early proposals to detect cosmic neutrinos
using an optical refractive effect \cite{Opher:1974, Lewis:1979mu},
which however does not give a net acceleration \cite{Cabibbo:1982bb, Langacker:1982ih}.
But there is another effect linear in Fermi's constant $G_F$ which was originally proposed by Stodolsky
\cite{Stodolsky:1974aq}, remarkably before the discovery of neutral
currents. It is due to the energy splitting of the two spin states of
the electrons of the detector material in the bath of cosmic neutrinos.
If there is an asymmetry between the densities of neutrinos and anti-neutrinos
in the CNB, this results in a net torque force on the test mass.
The acceleration of the test mass of the pendulum reads
\cite{Duda:2001hd}
\begin{equation}
 a_{G_F} = \frac{N_{AV}}{A \, m_{AV}} \frac{\Delta E}{\pi} \frac{\gamma}{R} \;,
\end{equation}
where $N_{AV}/(A  \, m_{AV})$ is
the number of nuclei in $1$~g test material.
$N_{AV} = 6.022 \cdot 10^{23}$ is Avogadro's constant, $A$
the number of nucleons in an atom and $m_{AV} = 1$~g is introduced
here for proper normalization.
$R$ is the radius of the test mass and $\gamma = M R^2/I$
is a geometrical factor related to the moment of inertia $I$ of the
detector with mass $M$.
Using the expression found in Ref.~\cite{Duda:2001hd} for the induced
energy splitting  $\Delta E$ of the electrons  we find
\begin{equation}
 a^R_{G_F} = \frac{N_{AV}}{A \, m_{AV}} \frac{2\sqrt{2}}{\pi} \,
   G_F \, \beta^\text{CMB}_\oplus \frac{\gamma}{R} \sum_{\alpha = e,\mu,\tau} (n_{\nu_\alpha} - n_{\bar{\nu}_\alpha}) \,  g^{\alpha}_A  \;,
\end{equation}
for relativistic neutrinos. For non-relativistic neutrinos
this could be at most one order of magnitude larger \cite{Duda:2001hd}.
Note that this effect only exists in the presence of a lepton asymmetry
in the CNB such that the number of neutrinos and anti-neutrinos do not cancel.
In the conventions of Ref.~\cite{Duda:2001hd}, $g_A^e = 0.5 = - g_A^{\mu,\tau}$.

The effect is larger for test masses with a small $A$ and
the probe has to be magnetized. This is most easily realized for
ferromagnets. The stable, elementary ferromagnet with the smallest
atomic number is the iron isotope with $A = 54$.
Furthermore we assume the test mass to be a massive sphere
with a radius of 1~cm ($\gamma = 0.4$) and we find
\begin{equation}
 \label{eq:GF}
 a^R_{G_F} \approx 4 \cdot 10^{-29} \,  \frac{n_{\bar\nu_\mu} - n_{ \nu_\mu} }{2 \, \bar{n}_\nu} \, \text{cm/s}^2 \;,
\end{equation}
where we have for simplicity taken the other two neutrino flavours to
be symmetric, $n_{\nu_{e,\tau}} = n_{\bar \nu_{e, \tau}}$.
With the expressions
above it is straight-forward to derive an estimate for more complicated
admixtures of flavours and neutrinos and anti-neutrinos.
Unfortunately, our estimate here is many orders of magnitude away from 
the benchmark sensitivity of $3 \cdot 10^{-18}$~cm/s$^2$,
cf.~\eqref{eq:sensitivity}.

\subsection{Scattering processes (\texorpdfstring{$\boldsymbol{G_F^2}$}{GF2} effect) \label{sec:scattering}}

Next we turn to the force due to the momentum transfer of CNB neutrinos
scattering off the target material, first discussed by
Opher \cite{Opher:1974} (for other early works, see, e.g.
\cite{Lewis:1979mu,  Shvartsman:1982sn}).
Since this force is proportional to $G_F^2$, one might expect this effect to be suppressed
compared to the magnetic effect discussed above. However, due to the macroscopic
wavelength of the low-energy CNB neutrinos, the cross section may be
enhanced not only by a nuclear coherence factor $\sim A^2$ but also by
a coherence factor $N_c$ from the scattering of multiple
nuclei~\cite{Opher:1974, Zeldovich:1981wf,Smith:1983jj}, see also
\cite{Freedman:1973yd}, such that this effect can be dominant.
It also does not require an asymmetry in the CNB.
The resulting acceleration of the test mass
can be written as \cite{Duda:2001hd} 
\begin{equation}
 a_{G_F^2} = \Phi_\nu \, \frac{N_{AV}}{A \, m_{AV}} \, N_c \, \sigma_{\nu-A} \, \langle \Delta p \rangle \;,
 \label{eq:GF2}
\end{equation}
where $\Phi_\nu = n_\nu \, p_\nu \, c^2/E_\nu$
is the neutrino flux with $E_\nu$ denoting the energy of the
CNB neutrinos and  $p_\nu$ the average relative momentum between
these neutrinos and the earth.
Further, $N_c$ denotes the coherence
enhancement factor, $\sigma_{\nu-A}$
the neutrino-nucleus cross-section containing the nuclear
enhancement factor and $\langle \Delta p \rangle$ is the average
momentum transfer from the scattered neutrinos.

The neutrino cross section at low energies (small recoil energies)
is \cite{Formaggio:2013kya}
\begin{equation}
 \sigma_{\nu-A} \approx \frac{G_F^2 }{4 \pi \, \hbar^4 \, c^4} \, (A-Z)^2 \, E_\nu^2 \;,
\end{equation}
with $E_\nu \approx m_\nu c^2$ for non-relativistic and $E_\nu \approx 3.15 \, k_B \, T_\nu$
for relativistic neutrinos.
In addition, the neutrinos will also scatter
off the electrons in the material. For $E_\nu \ll m_e c^2$ the cross section
to one electron is approximately \cite{Marciano:2003eq}
\begin{equation} 
 \sigma_{\nu_e - e} \approx \frac{7 \, G_F^2 }{4 \pi \, \hbar^4 \, c^4}  \, E_\nu^2
 \text{ and } \sigma_{\nu_{\mu, \tau} - e} = \frac{3}{7} \sigma_{\nu_e - e} \;.
\end{equation}
which is comparable to the nucleus cross section. However, contrary to the nucleus cross section, the electron cross section is sensitive to the flavour composition of the CNB and the effective momentum transfer from the electrons to the macroscopic target will depend on the details of the target material. We will hence omit this contribution in the following, noting however that including this effect may moderately increase the cross section.

The atomic coherence factor $N_c$ is given by the number of nuclei
within the de Broglie wavelength $\lambda_\nu = 2 \pi \hbar/p_\nu$ of the
neutrinos \cite{Duda:2001hd},
\begin{equation}
 N_c = \frac{N_{AV}}{A  \, m_{AV}} \, \rho \, \lambda_\nu^3 \;,
\end{equation}
where $\rho$ denotes the density of the test
mass at the end of the pendulum.
To maximize the coherence effect the test mass should ideally
have the same size as the de Broglie wavelength. Alternatively,
one could think of using foam-like \cite{Shvartsman:1982sn},
or laminated materials \cite{Smith:1983jj} or some embedding
of the detector material in a matrix material \cite{Smith:2003sy}.
For CNB neutrinos, the typical
de Broglie wavelength is ${\cal{O}}(0.1~\text{cm})$, leading easily to
$N_c \sim 10^{20}$.
So we can rewrite the acceleration as
\begin{equation}
 a_{G_F^2} = \frac{2  \, \pi^2 \, G_F^2}{\hbar \, c^2} \, n_\nu  \, \frac{N_{AV}^2 (A-Z)^2 \, \rho }{A^2  \, m_{AV}^2} \frac{\langle \Delta p \rangle E_\nu}{p_\nu^2}\;.
\end{equation}
Now we are left with $\langle \Delta p \rangle$, which can be
categorized roughly into three cases, see the discussion
in Sec.~\ref{sec:review}.
The first case are relativistic neutrinos $m_\nu \, c^2 \ll k_B T_\nu$
where
\begin{equation}
p_\nu^\text{(R)} \simeq 3.15 \, k_B T_\nu/c \,, \quad \langle \Delta p \rangle_\text{(R)} \simeq 3.15 \,  {\beta_{\oplus}^\text{CMB}}  \, k_B T_\nu/c \;,
\end{equation}
with the Boltzmann constant $k_B = 1.38 \cdot 10^{-16}$~g~cm$^2$~s$^{-2}$~K$^{-1}$
and the speed of light $c = 2.998 \cdot 10^{10}$~cm/s.
The factor $3.15$ arises from the thermal average over the Fermi-Dirac
distribution.
Here ${\beta_{\oplus}^\text{CMB}}$ denotes the velocity of the earth in the CNB frame.
If the earth was at rest in this frame, the neutrinos would arrive uniformly
from all directions and the average momentum transfer would vanish. The net
effect is thus proportional to the velocity of the earth (the pendulum) moving
through the neutrino bath.

Next we consider non-relativistic neutrinos ($m_\nu \, c^2 \gg k_B T_\nu$) which can be divided into two sub cases. First, we consider neutrinos which
do not cluster gravitationally.
Even though they are non-relativistic, their average
momentum is thus determined by the CNB temperature,
\begin{align}
 p_\nu^\text{(NR-NC)} \simeq 3.15 \, k_B T_\nu/c \,, 
  \quad \langle \Delta p \rangle_\text{(NR-NC)} \simeq 3.15 \,  {\beta_{\oplus}^\text{CMB}}  \, k_B T_\nu/c \;.
\end{align}
Note that in general, the relative momentum should be estimated as
\begin{equation}
(p_\nu)^\text{(NR-NC)} \simeq \text{max}\left\{ 3.15 \, k_B T_\nu \, , \; m_\nu \, \beta_{\oplus}^\text{CMB} \, c \right\}.
\end{equation}
However, since we are considering non-clustered neutrinos, their velocity must
be larger than the escape velocity. Since $v_\text{esc} \gtrsim \beta_{\oplus}^\text{CMB} c$,
the former term will always dominate.
On the other hand, the rest frame of clustered non-relativistic neutrinos
is the frame of the galaxy or (super) cluster. As a reference value for their
velocity dispersion we use $\beta_\text{vir}$, which also determines the relative velocity
of the earth in this frame, $v_\oplus \simeq \beta_\text{vir} \, c$,
\begin{equation}
   p_\nu^\text{(NR-C)} \simeq m_\nu \, \beta_\text{vir} \, c   \,, \quad  \langle \Delta p \rangle_\text{(NR-C)} \simeq  m_\nu \, \beta_\text{vir}\, c \;, 
\end{equation}
With this we find that the acceleration is given by
\begin{equation}
 a_{G_F^2} =  \frac{2  \, \pi^2 \, G_F^2}{\hbar \, c^2} \, n_\nu \, \frac{N_{AV}^2 (A-Z)^2 \, \rho }{A^2  \, m_{AV}^2}  \begin{cases}
   \beta_{\oplus}^\text{CMB} c  & \text{for (R)} \;, \\                              
   \frac{ m_\nu \, \beta_{\oplus}^\text{CMB} \, c^3}{3.15 \, k_B \, T_\nu} & \text{for (NR-NC)} \;, \\
   c/\beta_\text{vir} & \text{for (NR-C)} \;. \\
 \end{cases} 
\end{equation}

We can now plug in some numbers and compare them to our benchmark sensitivities.
As a target we choose lead which has a high density, $\rho \approx 11.34$~g/cm$^3$,
with $A = 208$ and $Z=82$~\footnote{
We choose here a material with a high neutron and mass density to 
increase the induced accelerations. Currently in gravitational wave experiments other
lighter materials are preferred to, e.g., reduce thermal noise. For pure silicon,
for instance, the acceleration would drop roughtly by a factor of seven.
}. 
In total we then obtain
\begin{equation}
a_{G_F^2} = \frac{n_\nu}{2\, \bar{n}_\nu}  \begin{cases}
   3 \cdot 10^{-33} \, \text{cm/s}^2  & \text{for (R)} \;, \\                               
   5 \cdot 10^{-31} \, (m_\nu/0.1~\text{eV/$c^2$}) \, \text{cm/s}^2  & \text{for (NR-NC)} \;, \\
   2 \cdot 10^{-27} \, (10^{-3} /\beta_\text{vir}) \, \text{cm/s}^2  & \text{for (NR-C)} \;, \\
\end{cases}
\end{equation}
Here we have normalized the neutrino density to the standard value of
$2 \, \bar n_\nu$.
We find here slightly different numbers than
Ref.~\cite{Duda:2001hd}. Apart from some improved approximations,
the main difference is the expression for $p_\nu$ employed
in the (NR-NC) case.

As discussed in section~\ref{sec:review}, various mechanisms
can (moderately) enhance these values. We stress that at least two neutrino generations
should be non-relativistic
nowadays, and that they are moreover at least partially clustered (at least
to the local super cluster) which is the more promising case for a potential discovery.
Nevertheless, these rates are many orders of magnitude below the benchmark sensitivities
quoted in Sec.~\ref{sec:pendulum}.

For completeness, let us estimate the lower bound on the induced acceleration,
taking into account all remaining uncertainties about the CNB. As can be seen from
the expressions above, the `worst case scenario' are light (normal ordered) Majorana neutrinos.
If the lightest neutrino is massless we find for the other two neutrino masses
$m_2 \approx 8.5 \cdot 10^{-3}$~eV/c$^2$
and  $m_3 \approx 5 \cdot 10^{-2}$~eV/c$^2$ which is much larger than their kinetic
thermal energy $3.15 \, k_B \, T_\nu \approx 5 \cdot 10^{-4}$~eV such that both of these species
can be considered to be non-relativistic. To be more precise their velocities are then
$v_2 \approx \sqrt{2E/m_2} \approx 0.2 \, c$ and
$v_3 \approx \sqrt{2E/m_3} \approx 0.1 \, c$ which is far above the local escape velocities.
Furthermore, the scattering cross section for Majorana neutrinos is suppressed by an
additional factor of $(v/c)^2$, see, e.g.,~\cite{Duda:2001hd}.
Combining this information we find that the minimal acceleration is of the order
\begin{equation}
  a_{G_F^2}^{\text{min}} \approx 8 \cdot 10^{-33} \, \text{cm/s}^2 \;,
\end{equation}
where we have summed over all three flavours. This is nearly six orders of magnitude
worse than the most optimistic scenario. However, 
once the neutrino mass scale and ordering is determined the uncertainty
on the induced acceleration will shrink significantly.
A determination of the Dirac / Majorana nature of neutrinos would furthermore significantly
reduce this uncertainty. Vice versa, if neutrino-less double beta decay experiments
remain inconclusive, the CNB could one day be a last resort to distinguish these
two possibilities.

As we have just seen the scattering effect can indeed be much smaller than the
$G_F$ effect, cf.~\eqref{eq:GF}, which makes it tempting to focus on the latter effect.
But this depends on the lepton asymmetries for which there is no
such clear theoretical prediction.

\subsection{Solar neutrinos and dark matter}

In this section we discuss two relevant competing processes which could also result in a
pendulum displacement with a similar frequency: solar neutrinos and particle dark matter (DM),
see also~\cite{Duda:2001hd}.

The acceleration due to solar neutrinos is given by Eq.~\eqref{eq:GF2} for
relativistic neutrinos, taking into account that the maximal momentum transfer
of solar neutrinos is simply given by $\langle \Delta p \rangle = E_\nu/c$
(since all solar neutrinos come from the same direction). The de Broglie wavelength of
the solar neutrinos is ${\cal O}(10^{-10})$~cm and hence smaller than
the typical atomic distances, implying that there will be no coherent
enhancement as for the CNB neutrinos ($N_c = 1$). With the solar neutrino flux
$\Phi_{\text{solar}-\nu} = 10^{11} \, \text{cm}^{-2} \, \text{s}^{-1}$, the scattering of
pp neutrinos ($E_\nu \simeq 0.3$~MeV) off a lead test mass yields
\begin{equation}
a_{\text{solar}-\nu} \approx 3 \cdot 10^{-26} \, \text{cm/s}^2 \,.
\end{equation}
This acceleration is larger than the CNB wind, however as we discuss in the following
section, the event rate is much smaller so that (in an earth based laboratory) the expected
signal would be clearly distinguishable. Moreover, in contrast to the CNB signal, this signal
will be correlated with the relative position of the sun.

Eq.~\eqref{eq:GF2} also applies to the acceleration induced by collisions
with cold dark matter particles $X$. In this case, the
corresponding flux is given by $\Phi_X = n_X \beta_X c$ and the momentum
transfer by $\langle \Delta p \rangle_X = m_X \beta_X c$ where $n_X$,
$\beta_X \, c$ and $m_X$ denote the number density, average velocity and
mass of the particles $X$. For dark matter masses $m_X \gtrsim 1$~GeV/$c^2$,
as expected in the WIMP scenario, the de Broglie wavelength is smaller
than $10^{-10}$~cm and we can set $N_c = 1$. Together, for a lead target
as considered above this yields
\begin{equation}
a_\text{DM} \approx 4 \cdot 10^{-30}  \left( \frac{(A - Z)^2}{76 \, A} \right) \left( \frac{\sigma_{X-N}}{10^{-46} \, \text{cm}^2}\right) \left( \frac{\rho_\text{dark(local)}}{10^{-24} \, \text{g}/\text{cm}^3}\right) \left( \frac{\beta_X}{10^{-3}} \right)^2 \text{cm/s}^2 \,,
\label{eq:DM}
\end{equation}
with $\rho_\text{dark(local)} = m_X n_X$ the local dark matter density,
implying that the benchmark sensitivity of Eq.~\eqref{eq:sensitivity_now}
corresponds to cross sections in this dark matter
range of $\sigma_{X -N} \gtrsim 2 \cdot 10^{-33}~\text{cm}^2$. 
In the prototypical WIMP mass range of GeV/$c^2$~$< m_X <$~100~GeV/$c^2$,
such a cross section is excluded by many orders of magnitude by current
direct detection searches, which find $\sigma_{X-N} \lesssim 10^{-46}~\text{cm}^2$
for spin-independent nucleon-DM interactions
for $m_X \approx 40 - 50$~GeV/$c^2$~\cite{Akerib:2016vxi,Tan:2016zwf}.

Given these strong constraints, sub-GeV DM candidates have recently 
received a lot of attention, see, e.g., \cite{Alexander:2016aln} and references
therein. In this mass range, the constraints from direct detection become 
irrelevant and the strongest bounds are derived from cosmology and astrophysical
considerations~\cite{Green:2017ybv}. For example, a thermal dark matter candidate
which is lighter than about 10~MeV/$c^2$ would decouple from the Standard Model
after the decoupling of the CNB, and hence generically perturb the standard
$T_\nu/T_0$-relation - which in turn is bounded by the $\Delta N_\text{eff}$
measurements in the CMB~\cite{Ho:2012ug,Boehm:2013jpa}. Further constraints 
arise from the relic abundances of the elements produced in Big Bang
Nucleosynthesis~\cite{Serpico:2004nm,Nollett:2013pwa,Nollett:2014lwa} 
and from bounds on CMB distortions~\cite{Finkbeiner:2011dx,Lopez-Honorez:2013lcm}.
Assuming standard cosmology, these constraints exclude wide mass-ranges of
sub-MeV thermal relics~\cite{Green:2017ybv},  subject however to assumptions on, e.g., the annihilation
channels, the nature of the mediator fields and the production mechanism.

In contrast, it is interesting to note that the setup we propose here could function
as a fairly model-independent direct DM detector.
In the sub-MeV mass range, our proposal gains ground in a two-fold way: Firstly, the
DM number density increases as $1/m_X$ implying that DM acts more like a constant
`wind' (as in the CNB case) instead of separate individual events even for lower
cross sections. Secondly, and more importantly, the atomic enhancement factor
$N_c$ begins to rapidly grow for $m_X \lesssim$~MeV/$c^2$, reaching $N_c \sim 10^{9}$
for a lead test mass at $m_X = 3.3$~keV/$c^2$ (which corresponds to the lower bound on the DM mass from
structure formation for a thermal relic~\cite{Viel:2013apy}).
In this mass range our setup with the benchmark sensitivity of Eq.~\eqref{eq:sensitivity_now}
could thus allow to probe DM - Nucleon cross sections
down to $\sigma_{X -N} \simeq 10^{-42}~\text{cm}^2$. Note that even lighter DM masses are
theoretically viable if the DM particle has a suitable non-thermal history and hence its
contribution to washout during structure formation is suppressed.
These numbers should be compared with other very recent proposals to directly
measure DM in this mass range, see, e.g., Refs.~\cite{Essig:2011nj,Graham:2012su,Essig:2015cda,Hochberg:2016ajh,Schutz:2016tid,
Derenzo:2016fse,Hochberg:2016sqx,Essig:2016crl,Knapen:2016cue}.

A dark matter signal might be disentangled from a CNB signal through the
phase of the annual modulation~\cite{Safdi:2014rza}. The fractional
modulation for light, unbound neutrinos is expected to peak in fall,
whereas the peak for bound particles (such as DM) is expected in spring,
see Fig.~2 of~\cite{Safdi:2014rza}. This is of course only possible if
the CNB neutrinos are mainly unclustered. Note that the expected signal
from bosonic sub-eV dark matter as discussed in~\cite{Graham:2015ifn}
oscillates with a frequency of $m_X/(2 \pi \hbar) \sim m_X/(0.05~\text{eV}/c^2) \cdot 10^{13}$~Hz.
For $m_X \gtrsim 10^{-19}~\text{eV}/c^2$ this is much faster than the 1/day
frequency of the signals discussed here.


\subsection{Cosmic wind vs.\ cosmic nudges \label{sec:rates}}

A question which has not explicitly been addressed in the literature to our knowledge
is, if the CNB really acts like a wind or if it is more realistically a series
of feeble nudges on the test mass. To understand this issue better let us first
have a look at the event rates of CNB neutrinos, focusing for simplicity here
only on the $G_F^2$ effect. From
\begin{equation}
 R = \frac{a_{G_F^2}}{\langle \Delta p \rangle} = \Phi_\nu \, \frac{N_{AV}}{A \, m_{AV}} \, N_c \, \sigma_{\nu-A}  \;,
\end{equation}
we obtain
\begin{align}
 R_\text{(R)} &\approx 1 \cdot 10^{-4} \, \frac{n_\nu}{2 \, \bar{n}_\nu} \, \text{g}^{-1} \, \text{s}^{-1} \;, \\
 R_{\text{(NR-NC)}}  &\approx  0.02 \, \frac{n_\nu}{2 \, \bar n_\nu} \, 
    \frac{m_\nu}{0.1~\text{eV/$c^2$}} ~\text{g}^{-1} \, \text{s}^{-1} \;,\\
 R_{\text{(NR-C)}}  &\approx 0.4 \, \frac{n_\nu}{2 \, \bar n_\nu}  \, \frac{0.1~\text{eV/$c^2$}}{m_\nu}
  \left( \frac{10^{-3}}{\beta_\text{vir}} \right)^2 ~\text{g}^{-1} \, \text{s}^{-1} \;,
\end{align}
using the above results and with lead as detector material. For a total test mass
of about 100~kg (arranged properly to fully exploit the atomic coherence factor) the expected
frequency of events is in all three cases larger than the oscillation frequency of the
pendulum ($f = 1/(2 \, \pi) \sqrt{g/l} \approx 0.5$~Hz), and it is indeed justified to
speak of a neutrino `wind'.
Interestingly one could also follow another approach here. It is difficult
to keep the interferometers stable on the time scale of a day. Instead,
one can choose the material, the size of the detector and the length of the
pendulum in such a way that the signal appears as noise in the frequency band
where the detector is most sensitive (i.e., $R \sim 100$~Hz for LIGO).
The daily/annual modulation as well as the preferred average direction
of the signal would then result in a fluctuation of the noise which might be
easier identified than the very low-frequency signal discussed above
as we also have pointed out already in section~\ref{sec:pendulum}.

Interestingly, the event rate for solar neutrinos is very low
\begin{equation}
 R_{\text{solar}-\nu} \approx 2 \cdot 10^{-9} \, \text{g}^{-1} \, \text{s}^{-1} \,
\end{equation}
which might seem surprising due to the very large flux and the much higher
nucleus cross section. But in this case the missing atomic coherence factor
really makes a big difference. Consequently, solar neutrinos will register in
our setup as a series of individual nudges. Disentangling these from
other background events seems challenging but again the
directional information can help.

For WIMP-like cold dark matter the rate is given by
\begin{equation}
  R_{\text{DM}} \approx 8 \cdot 10^{-3} \left( \frac{100~\text{GeV/$c^2$}}{m_X} \right) \left( \frac{\sigma_{X-N}}{10^{-33} \, \text{cm}^2}\right) \left( \frac{\rho_\text{dark(local)}}{10^{-24} \, \text{g}/\text{cm}^3}\right) \left( \frac{\beta_X}{10^{-3}} \right) \, \text{g}^{-1} \, \text{s}^{-1} \;, 
\end{equation}
normalized to our expected sensitivity for the cross-section for 100~GeV/$c^2$
dark matter as discussed above. 
This rate is extremely small once one inserts the direct detection constraints,
$\sigma_{X-N} \lesssim 10^{-46} \, \text{cm}^2$, as  expected compared to
plausible rates in dark matter direct searches.
However, this picture drastically changes
considering light dark matter due to the atomic coherence factor. 
For $m_X \lesssim 1$~MeV/$c^2$,
\begin{equation}
 R_{\text{light DM}} \approx  4 \cdot 10^5 \, \left( \frac{3.3 \text{ keV/$c^2$}}{m_X} \right)^4 \left( \frac{\sigma_{X-N}}{10^{-42} \, \text{cm}^2}\right) \left( \frac{\rho_\text{dark(local)}}{10^{-24} \, \text{g}/\text{cm}^3}\right) \left( \frac{\beta_X}{10^{-3}} \right) \, \text{g}^{-1} \, \text{s}^{-1} \;, 
\end{equation}
which is now normalized to our expected
sensitivity for the cross-section in this mass region.
Here we find large interaction rates even for the low
cross-sections considered, which is again mainly due to the atomic
coherence factor.


\section{Summary and conclusions}

Two of the most outstanding achievements in cosmology in recent years were
the precise measurement of the Cosmic Microwave Background and the detection
of gravitational waves. These remarkable discoveries raise the appetite for more. In
particular, in this paper we address the question if the impressive laser
interferometer technology used in gravitational wave detectors can be used
to hunt for an echo of the Big Bang generated much earlier than the CMB:
the Cosmic Neutrino Background.

Unfortunately, this does not seem to be feasible with the current technology.
We have briefly sketched a setup based on an ordinary pendulum which is deflected
by the cosmic neutrino wind. Using current laser interferometers to determine
the position of the pendulum and assuming a very high stability of the
experiment on the time-scale of a day it might be possible to measure
accelerations down to $10^{-16}$~cm/s$^2$, which would already be
an improvement
compared to current torsion balance experiments.
Although this sensitivity is already extremely remarkable it is still far
away from a potential signal. The most optimistic case for this kind of
experiments is when the relic neutrinos are non-relativistic nowadays and
cluster in our galaxy. This could lead to accelerations of the order of
$10^{-27}$~cm/s$^2$. Eleven orders of magnitude below what we estimated
might be ideally achieved with current technology. In addition, the low
frequency of the CNB signal poses a further serious challenge. A possibility
to address this point is by tuning the setup to the high frequency component
of the CNB signal, governed by the neutrino interaction rate with the test
mass. However, in summary,
these results suggest that a mechanical force might not be the most
encouraging way to discover the CNB. More promising for the discovery
at the moment seems an experiment
where cosmic neutrinos are caught with inverse beta decay giving rise to a
characteristic peak in the beta spectrum, see the recent PTOLEMY proposal
\cite{Betts:2013uya}. Such an experiment nevertheless comes with one big
disadvantage: It is not immediately sensitive to the directional dependence
of the CNB signal.

In the future, one might still want to consider an experiment along the lines
discussed here. In fact, the sensitivity could be tremendously
improved by putting the setup in a micro-$g$ environment, which could be
achieved either by going to space or by compensating the gravitational force
by an electromagnetic force here on earth in a laboratory. Another possibility,
motivated by the remarkable results of the recent LISA Pathfinder mission,
could be a setup based on free falling test masses in space with different
CNB cross sections.

A setup along the lines proposed here could moreover also serve as a
dark matter detector for sub-MeV DM particles. In particular in the low mass region
of a few keV, remarkable sensitivities to the DM nucleon cross section may be reached
even with current technology. For example, for DM particles close to the thermal limit,
$m_X = 3.3$~keV/$c^2$, we demonstrate how cross sections down to
$\sigma_{X-N} \simeq  10^{-42}~\text{cm}^2$  could be probed assuming a sufficiently
high stability of the experimental setup.
Expected developments in interferometer technology and the possibility of micro gravity
environments have the potential to significantly improve this number.

\section*{Acknowledgements}

The authors wish to thank M.~Barsuglia and E.~Chassande-Mottin for helpful
discussions on the experimental setup of the LIGO detector, as well as S.~Knapen
and T.~Lin for valuable comments on the status of light dark matter searches.
Moreover, we wish to thank G.~Gelmini for helpful comments and clarifications
about the momentum distributions of CNB neutrinos and Xun-Jie Xu
for pointing out the effect of the electrons in the detector material.
M.\,S.\ would like to thank A.~Meroni for pointing out \cite{Safdi:2014rza} to him. 
V.\,D.\ acknowledges financial support from the UnivEarthS Labex program at
Sorbonne Paris Cit\'e (ANR-10-LABX-0023 and ANR-11-IDEX-0005-02) and
the Paris Centre for Cosmological Physics.
V.\,D.\ would like to thank UC Los Angeles and the Berkeley Center for Theoretical Physics for kind hospitality during the final stages of this work.

\end{document}